\documentstyle[psfig,frontieres,art10]{article} 
\begin{document} 
\heading{%
%
Searching data for periodic signals  
%
}  
\par\medskip\noindent 
\author{
Andrzej Kr\'olak$^{1,2}$ 
}
\address{Max-Planck-Institute for Gravitational Physics,
Albert-Einstein-Institute, 
Schlaatzweg 1, 14473 Postsdam, Germany. 
} 
\address{%
Institute of Mathematics,
Polish Academy of Sciences, \'Sniadeckich 8, 00-950 Warsaw, Poland
} 
\begin{abstract} 
We present two statistical tests for periodicities in the time series.
We apply the two tests to the data taken 
from Glasgow prototype interferometer 
in March 1996. We find that the data contain several 
very narrow spectral features. 
We investigate whether these features can be confused 
with gravitational wave signals from pulsars.
\end{abstract} 
\section{Introduction}
The work presented here was motivated by an analysis of Gareth Jones 
\cite{J1} of the data taken from Glasgow prototype in 1996. 
His visual inspection of the periodogram 
of the data revealed presence of 3 very narrow (1 bin wide) 
significant spectral features.

\section{Statistical tests for periodicities in the data using 
the discrete Fourier transform} 
A standard method to search the time series for periodic signals
is to perform the Fourier transform (FT) of the series and examine
the modulus of FT for significant values.
Let $x_n$ be a real-valued discrete time random process 
given at equally spaced intervals $\Delta t$ so that the {\em sampling frequency} 
$f_s$ is equal to $1/\Delta t$. Let the number of samples of $x_n$ be $N$ 
and let us assume for simplicity that $N$ is even.  
The periodogram $P(f)$, $f \geq 0$, of $x_n$ is defined by
\begin{equation}
P(f) = \frac{1}{N}\left|\sum^{N-1}_{n=0}x_{n+1}
\exp^{-i 2\pi\frac{f}{f_s} n}\right|^2.
\end{equation}
At {\em Fourier frequencies} $f_k = \frac{k}{N} f_s$, $k = 0,1,...,N/2$
the quantity in the modulus is the {\em discrete Fourier transform} (DFT)
of the time series (for non-negative frequencies) and can effectively 
be evaluated by means of the {\em fast Fourier transform} (FFT) algorithm.
For the case when $x_n$ are uncorrelated and drawn from a Gaussian
distribution with zero mean and variance $\sigma^2$ and consquently
that random variables $P(f_k)$ are exponentially distributed and independent,
Fisher, in a celebrated paper \cite{F1}, derived a mathematically 
exact test for the presence of a periodic signal in the data based on 
the statistics
\begin{equation}
g = \frac{\max_{1 \leq k \leq N/2-1}[P(f_k)]}{\sum^{N/2-1}_{j=1}P(f_j)}, 
\end{equation}
where $\max_{1 \leq k \leq N/2-1}$ means maximum taken over 
the values of the periodogram evaluated
at Fourier frequencies for $k = 1,...,N/2-1$. 
Fisher's test is the most powerful test against simple periodicities
i.e., where the alternative hypothesis is that there exists a periodicity
at only one Fourier frequency. Usually there maybe many periodic signals
in the data and the number of them may be unknown.
For this case Siegel \cite{S1} proposed a test based on large values of the
periodogram with statistics
\begin{equation}
T = \sum^{N/2-1}_{k=1}
\left(\frac{P(f_k)}{\sum^{N/2-1}_{j=1}P(f_j)} - \lambda g_o\right)_{+},
\end{equation} 
where $g_o$ is the critical value for Fisher's statistics, 
$\lambda$ is a parameter such that $0 < \lambda \leq 1$, 
and subscript $+$ denotes the positive part. 
When $\lambda = 1$ Siegel's test $T > 0$ 
is equivalent to Fisher's test.
Siegel derived exact probability distribution for his statistics. 
By means of the Monte Carlo simulations he found that for $\lambda = 0.6$
his test was only slightly less powerful than Fisher's test when one
periodic signal is present in the data but it was substantially more
powerful when 2 or 3 periodic signals were present.

In practice none of the assumption about the time series required for
Fisher's and Siegel's test are met. The time series may consist of
non-Gaussian correlated random variables and moreover the time series
may be non-stationary.
For stationary processes (not necessarily Gaussian) with continuous spectral density
it can be shown (under fairly mild conditions) that asymptotically
(i.e. as $N \rightarrow \infty$) periodogram values
are independent and exponentially distributed 
with probability density function (pdf) $p$ given by
\begin{equation}
p[P(f)] = \frac{\exp^{-\frac{P(f)}{S(f)}}}{S(f)},
\end{equation}
where $S(f)$ is two-sided spectral density 
function \cite{PW}. The main difficulty in using the above pdf is that usually
the spectral density is unknown and has to be estimated
from the data itself.
We can however obtain an approximate test as follows.
Take $L$ blocks of $R$ consequtive values of periodogram evaluated
at $M = L \times R$ Fourier frequencies.
Consider the following statistics for each block l.
\begin{equation}
g'_k = \frac{P(f_k)/S(f_k)}{\frac{1}{R}\sum^{lR}_{j=(l-1)R+1}P(f_j)/S(f_j)}.
\end{equation} 
Asymptotically $\max[g'_k]$ has the same distribution as Fisher's statistics
with $R$ degrees of freedom. 
One may assume that over a certain bandwidth $B$ of $R$ Fourier bins 
(i.e. $B = \frac{R}{N}f_s$) the spectral density $S(f_k)$ 
changes very little and can be replaced by a constant value. 
Then $S(f_k)$ cancels out in the above formula and $g'_k$ can be approximated by
\begin{equation}
g_k = \frac{P(f_k)}{\frac{1}{R}\sum^{lR}_{j=(l-1)R+1}P(f_j)}.
\end{equation} 
Therefore we propose the following test statistics $g_A$ and
$T_A$ for simple and compound periodicities respectively 
\begin{equation}
g_A = \max_{[1 \leq l \leq L]} \{\max_{[(l-1)R + 1  \leq k \leq lR}[g_k]\}
\label{GA}
\end{equation}
\begin{equation}
T_A = \frac{1}{M}\sum_{l=1}^L\sum_{k=(l-1)R+1}^{lR}(g_k - \lambda g_o), 
\label{TA}
\end{equation}
where $g_o$ is the critical value of Fisher's statistics for $M$
points. For $\lambda = 1$ the test $T_A > 0$ is equivalent to the test
test based on statistics $g_A$.
Asymptotically normalized periodogram values $g_k$ 
for different blocks are independent
random variables and using this fact one can calculate the probability 
distribution for $g_A$ and $T_A$. For $g_A$ the critical values are
given by
\begin{equation}
g_{oA} = R \{1 - [(1 - (1 - \alpha)^{R/M})/R]^{1/(R-1)}\}.
\label{goA}
\end{equation}
The above formula means that for $L=M/R$ blocks of $R$ points each
probability of statistics $g_A$ exceeding threshold the $g_{oA}$ 
in one or more bins out of the total $M$ bins
when the data is only noise is $\alpha$. 
In radar terminology $g_{oA}$ is called the {\em false alarm probability}.
For $M < 2^{24}$, $R > 2^7$, $\alpha < 0.01$ the critical values 
$g_{oA}$ can be approximated by $g_{oA}^a=-\log(\alpha/M)$ 
within an error of 7.5\%.
In turn $g_{oA}^a$ approximates the exact critical values for Fisher's statistics
with $M$ degrees of freedom 
for $M > 4\times 10^6$ and $\alpha < 0.01$ within 0.5\%.
Approximate critical values for statistics $T_A$ can be calculated 
from an asymptotic distribution for Siegel's statistics for $R$ points 
which is non-central $\chi^2$ 
distribution with zero degrees of freedom \cite{S2} and from the fact 
that convolution of non-central $\chi^2$  distributions is a again $\chi^2$.
Critical values can also be calculated to a reasonable approximation 
just from a non-cental $\chi^2$ distribution with zero degrees of freedom 
for $M$ points.

\section{Glasgow data}
We have applied the statistical tests described in Section 1 
to the data taken from the prototype interferometric detector
in Glasgow. This data was taken on 6th of March 1996 from
21:00:00 U.T. to 22:22:44 U.T.
The data consisted of 19857408 samples taken at 1/4 ms intervals
and quantized with a 12 bit analogue-to-digital converter with a dynamic
range from -10 to 10 Volts.

From time to time the detector was out of lock and the level of the noise
was very high. Even when the detector was in lock standard deviations 
for short blocks of data of $2^8$ to $2^{12}$ points varied 
showing that the data was not stationary. Calculation of skewness 
and kurtosis for short blocks of data revealed that data tended to have a
longer tail for negative values than for positive ones and that 
its distribution tends to be flatter with respect 
to normal distribution showing non-Gaussian behaviour of the data.
Applications of the standard spectral estimation techniques (Welch overlap
method with Hanning window and Thomson multitaper method) showed
that over the frequency range of 400Hz to 1.2KHz the spectral density 
consists of a reasonably flat part superposed 
with many narrow spectral features (see Figure 1).
\begin{figure} 
\centerline{\vbox{ 
\psfig{figure=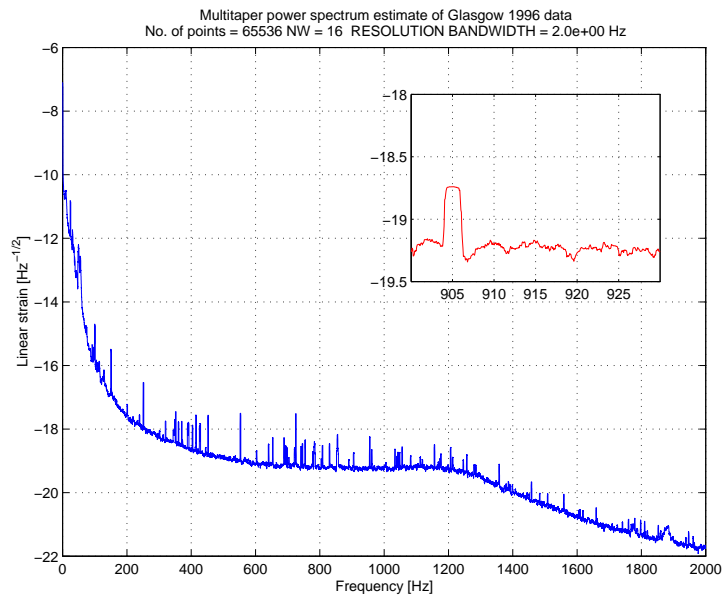,height=7.9cm} 
}} 
\caption[]{An estimate of two-sided spectral density of 1996 
Glasgow data. The spectral feature shown in the insert is due to
harmonics of the mains frequency.  
} 
\end{figure} 
The flat part corresponds to linear one-sided spectral density 
of around $10^{-19}$Hz$^{-1/2}$.

\section{Data preparation}
We have divided the data into blocks of $2^8$ points.
We have singled out blocks of 'bad' data by the following
criterion. We have selected those blocks in which the maximum 
of the absolute values of the data in the block exceeded 8.5 Volts. 
We have then defined the {\em window function} $W_n$, $1 \leq n \leq N$ 
as $0$ for each $n$ such that data sample $x_n$ is in 
the selected block of 'bad' data and $1$ otherwise.
We have also normalized the data set as follows.
In each block of data we have subtracted from every point the
block mean and divided by the block standard deviation. We have then multiplied
the resulting sequence by the window function $W_n$.
A similar procedure was applied in an analysis of 100 hours Garching data
by Niebauer et al. \cite{Nea}.
Dividing the data by the block mean improves the signal-to-noise ratio (SNR)
because periods of low noise make the highest contributions to overall SNR. 
Also the normalization reduces the slow variation of the mean
and the variance of the noise 
thus removing some non-stationarity from the data. 
  
\section{Results of the tests for periodicity}
We have calculated DFT of the whole data set (using the FFT algorithm)
and we have analysed the periodogram for periodicities 
in the frequency range from around 450Hz to 1250Hz i.e. around $4\times10^6$
Fourier bins altogether. We have divided DFT into blocks of length
$R = 2^7$ bins. We have chosen a very high significance level
$\alpha$ of $10^{-6}$. We have applied a test based on the statistics
$g_A$ in the following way: we have calculated the threshold form the
Eq.~(\ref{goA}) for $\alpha = 10^{-6}$ and we have registered all
the values of the normalized periodoram $g_k$ that crossed this threshold.
The test yielded 14 significant events.  
The first 6 of them are shown in Figure 2.
\begin{figure} 
\centerline{\vbox{ 
\psfig{figure=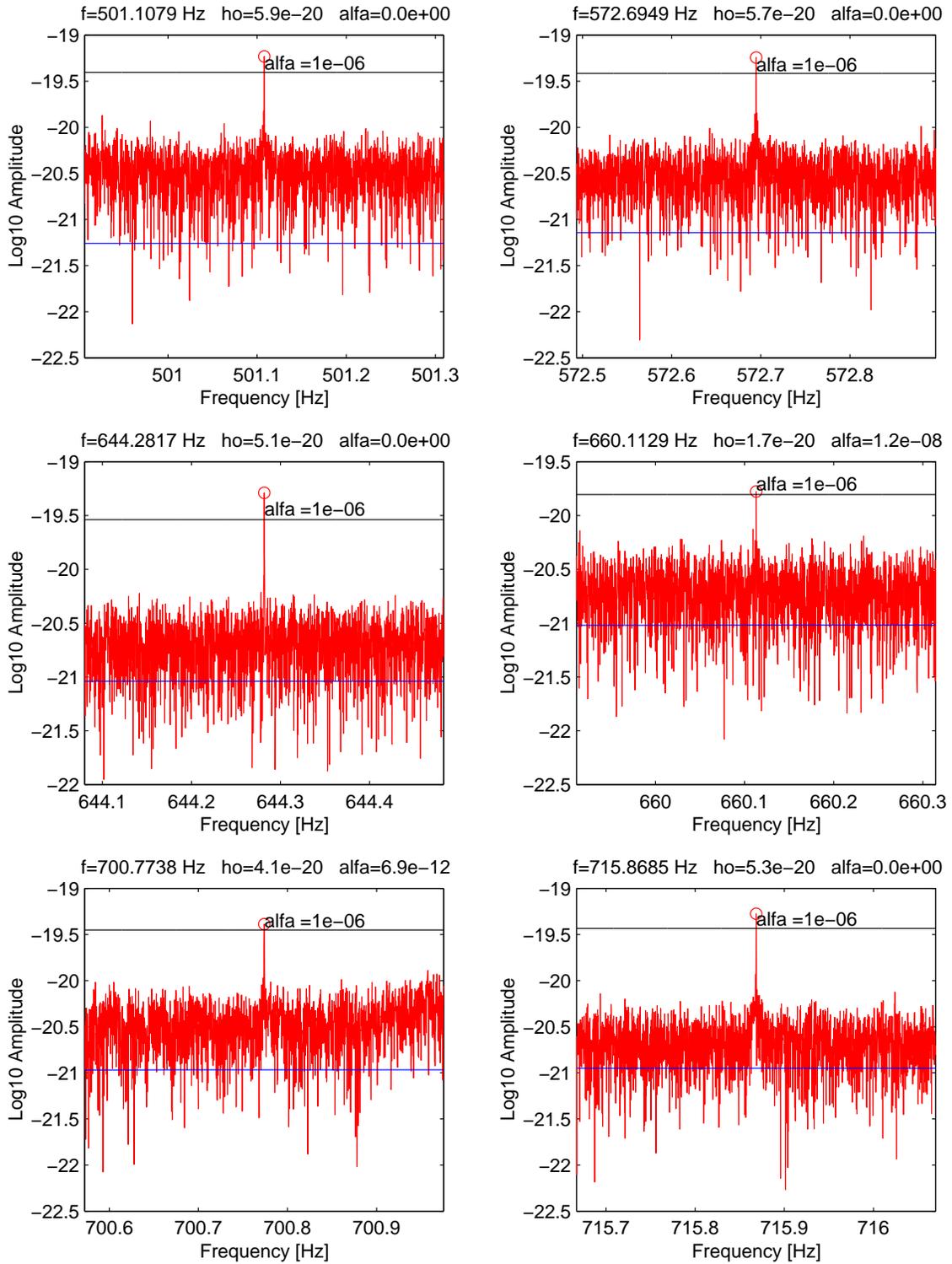,height=20.0cm} 
}} 
\caption[]{Narrow spectral features in the 1996 Glasgow data.  
} 
\end{figure} 
They were all narrow lines, 1 to 2 bins wide. We have also found
that all these lines were harmonics of the following set of frequencies:
$h_1=60.0103$Hz, $h_2=70.0774$Hz, $h_3=71.5869$Hz.
On top of each frame in Figure 1 we have given the frequency $f$ of the
detected spectral feature, the dimensionless amplitude $h_o$, 
the significance level $\alpha$ of the event calculated from Eq.~(\ref{goA}) 
($\alpha=0$ means that it is smaller than machine accuracy 
$\sim 2.2 \times 10^{-16}$).
If the spectral feature corresponds to a monochromatic signal 
form outside the detector then $h_o$ is the maximum-likelihood estimator
of its amplitude.
The upper line is the threshold corresponding to $10^{-6}$ significance 
level. The lower line is what we call "pulsar line". 
For the detector located in Glasgow it corresponds
to the maximum amplitude of the gravitational wave of a pulsar 
at twice the pulsar spin frequency 
assuming ellipticity of $10^{-4}$, distance 40pc from the Earth, and 
moment of inertia of $10^{45}$gcm$^2$  w.r.t. the rotation axis.
We consider this as the strongest pulsar signal possible with
our current understanding of pulsar distribution in the galaxy and
their physics.
The results of Siegel's test revealed 27 events: 7 more harmonics
of the frequencies given above. One harmonic of frequency $h_4=72.8174$Hz
(only one more harmonic of that frequency was found 
for much lower significance level of $5\times 10^{-2}$ with 
$g_A$ test), 2 narrow features riding on top of wide spectral features
of bandwidth $~0.1$Hz, and three narrow, 1 bin wide lines
of frequencies 510.4761Hz, 511.1870Hz, and 1210.5961Hz that could
not be related to any harmonics.
The amplitudes of the last three features were $35, 30, 26$ times above
the pulsar line respectively.
Two of the frequencies found by Jones \cite{J1} where 8th and 
11th harmonics of the frequency $h_1$ given above, 
the third one $f_{G3} = 675.0879$Hz
was none of the frequencies reported above. Nevertheless we confirmed its
existence in the data with a very low significance $\alpha=0.44$.

Comparision of the results of the two tests shows that the 
test based on the statistics $T_A$ is considerably more powerful in detecting
periodicities in the spectrum than the test based on the statistics $g_A$.
We have repeated the above analysis for various length of the 
data blocks and another criterion of tagging the bad data
based on the magnitude of the variance in the block and also for various
length $R$ of DFT blocks and the results of the above analyis have not changed
substancially.
  
\section{Conclusions}
Our conclusion is that none of the spectral features detected by us
could be confused with pulsar signals. Firstly we would not expect
a gravitational wave from a pulsar to show in the Fourier domain as
a series of harmonics. We can expect significant power around once and twice
the pulsar spin frequency with harmonics of a much smaller amplitude.
Secondly the amplitudes of all the spectral features including
narrow single frequencies are much higher than for any possible gravitaional
wave from a pulsar.
Finally we point out that given only a finite number of samples
of the data and no further information it is impossible 
to distinguish strict periodic components from peaks of arbitrary small width
in the continuous spectrum \cite{Pr}. 
The spectral lines that we detected are due to a periodic
deterministic signal in the data if we know that the  maximum
width of the spectral features in the continuous part of the
spectrum is greater than $R$ Fourier bins.

\acknowledgements{I would like to thank Albert Einstein Institute,
Max Planck Institute for Gravitational Physics for hospitality.
I am grateful to Bernard F. Schutz for many important suggestions, 
Morag Casey for helpful discussions, and Jurek Usowicz for help in
numerical work.} 

%
\vfill 
\end{document}